\documentclass[IJQI,twocolumn,showpacs,superscriptaddress,preprintnumbers,amsmath,amssymb,floatfix]{revtex4}

\usepackage{graphicx}
\usepackage{dcolumn}
\usepackage{bm}

\begin{document}


\title{Quantum informatics with plasmonic metamaterials}

\author{Ali A. Kamli}
\affiliation{Department of Physics, University of Jazan, Jazan, P O Box 114, Saudi Arabia}
\author{Sergey A. Moiseev}
\affiliation{Kazan Physical-technical Institute of Russian Academy of Sciences, 10/7 Sibirsky Trakt, Kazan, 420029, Russia}
\author{Barry C. Sanders}
\affiliation{Institute for Quantum Information Science, University of Calgary,
	Calgary, Alberta, Canada T2N 1N4}

\date{\today}


\begin{abstract}
Surface polaritons at a meta-material interface are proposed as qubits. The surface-polariton fields are shown to have low losses, subwavelength confinement and can demonstrate very small modal volume. These important properties are used to demonstatre interesting applications in quantum information, i.e., coherent control of weak fields and large Kerr nonlinearity at the low photon level.
\end{abstract}

\pacs{42.50.Gy, 42.25.Bs, 78.20.Ci}

\maketitle

\section{Introduction}
\label{sec:introduction}

In quantum information, quantum bits, or qubits, are the fundamental entity for encoding information~\cite{Bou01}. Qubits are two state quantum systems examples of
such systems include photon vertical and horizontal polarization states, spin states, energy levels of an atom, etc~\cite{Nel07,KMN+07}. The interaction of light with atoms and the ability of coherent quantum control of these interactions is a basic problem of quantum information science. For instance, quantum memory requires high coupling and reversible  dynamics in the interaction between the photons and atoms in order to provide high efficient storage and retrieval of light. Furthermore the carriers of quantum information, qubits, have to be processed and manipulated through unitary quantum gates. The operation of some of these gates, e.g control phase shift operation (CZ gate), requires media with large Kerr nonlinearity that conventional media can not provide. Coherent control of light field was a subject of numerous investigations and considerable progress has been achieved in the last decades to develop many techniques including electromagnetically induced transparency (EIT)~\cite{HH99},
coherent population oscillation, and hole burning, for more extensive reviews the reader is referred to~\cite{Mil05,FIM05}.

Following our recent work ~\cite{KMS08}, we suggest in this review article surface polaritons as qubits for information processing. Surface polaritons (SP) are highly confined electromagnetic excitations at the interface of two media~\cite{AM82,Mai07}.

Strong spatial confinement of SP leads to a huge enhancement of the electromagnetic field near the media surface.
In conventional media (with permeability $\mu=1$) only the TM polarized electric SP modes can exist. Promising properties of SP field can be realized with the advent of artificially fabricated negative index metamaterials (NIMM) with both the permittivity $\varepsilon<0$ and permeability $\mu<0$ ~\cite{Ves67,Pen00,Sha07,EZ06}: 
(1) both types (TM and TE mode) of polarizations exist. For quantum information applications, supporting both modes would be highly desirable 
in allowing polarization qubits,
(2) low losses at frequency range that can be well controlled using external parameters,
(3) highly confined field amplitude. This is useful for enhancing the atom-field coupling which is important for coherent control and for increasing optical depth required in quantum memory, and
(4) large Kerr nonlinearity and order of $\pi$ phase shifts are possible to achieve in such media.

Below we demonstrate these properties of SP fields on the interface of dielectric and negative index metamaterial (NIMM). First we introduce the concept of the surface polaritons and photonics and show that these suggested qubits are capable of generating many interesting applications. In particular we demonstrate control of slow SP modes and the possibility of nanoscale SP based quantum memory devices~\cite{FL02,MRKZS01,KP00,Moi07,KTGNKC06,ALSM06}. Coherent control of SP qubits is achieved using EIT approach which provides a considerable longitudinal compression of the SP modes and enhancement of the interaction time with atoms near the dielectic/NIMM interface. 
Double EIT ( DEIT)  is a promising tool to obtain large Kerr nonlinearity for free propagating single photon fields~\cite{SI96,HH00,LI01,PK02}.
Here we demonstrate large Kerr nonlinearity and cross phase shifts for interacting weak SP fields.
A giant cross-phase modulation between the two SPs is achieved in a low-loss, sub-wavelength confinement regime.
A mutual $\pi$ phase shift between the two SP pulses is attainable for the  fields with a mean photon number of one, thereby opening the prospect of deterministic single-photon quantum logic two qubit gates for quantum computing~\cite{KMN+07}.

\section{Low loss surface polariton qubits}
\label{sec:dispersion}

\begin{figure}[htp]
  \centering
  \includegraphics[width=80mm]{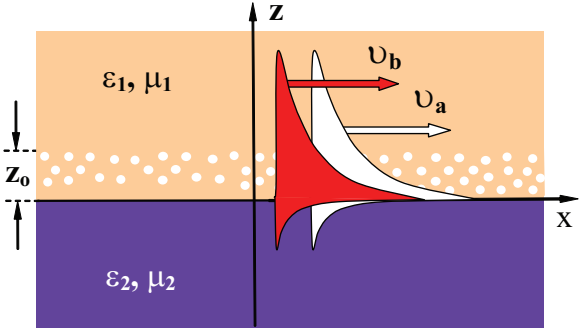}
  \caption{Surface polaritons are excited at a planar interface,
  between a dielectric in the upper half-space~$z>0$ and a metamaterial occupying the lower half-spaces $z<0$.
  Propagation is in the positive $x$-direction.The dots above the interface represent multi-level atoms.}
\label{fig:scheme}
\end{figure}

A meta-material interface Fig.~\ref{fig:scheme}, supports surface polaritons with both transverse magnetic (TM) and transverse electric (TE) polarization modes and these polarization degrees of freedom allow new type of qubits, i.e. surface polariton qubits. In the following we shall explore some of their basic physical properties. The surface polaritons (SP) can be excited at the interface (located at $z=0$) of two media using one of the standard techniques ~\cite{Mai07}.
The first medium is assumed dielectric with constant electric permittivity
$\varepsilon_0\varepsilon_1$ and magnetic permeability $\mu_0\mu_1$,
occupies the half space $z>0$, and the second medium is assumed to be a NIMM medium which occupies the half space $z<0$, with electric permittivity $\varepsilon_0\varepsilon_2(\omega)$ and magnetic permeability $\mu_0\mu_2(\omega)$.

The SP fields propagate with frequency~$\omega$ along the $x$-direction parallel to the interface with complex wave number
(for TM polarized SP fields) 
\begin{equation}
\label{eq:TM-SP-existence-b}
	K_\parallel=k_\parallel+\text{i}\kappa=\frac{\omega}{c}
		\sqrt{\varepsilon_1\varepsilon_2\frac{\mu_1\varepsilon_2-\mu_2\varepsilon_1}
			{\varepsilon_2^2-\varepsilon_1^2}}.
\end{equation}
 The wave numbers normal to the interface are related as;
\begin{equation}
\label{eq:TM-SP-existence-a}
	k_1\varepsilon_2+k_2\varepsilon_1=0,
\end{equation}
where $k_j^2=K_\parallel^2-\omega^2\varepsilon_j\mu_j/c^2$
is the $z$-component of the wave-vector normal to interface ($j =1,2$), with the indices~1 and~2 referring to the two media.
With similar expressions for TE polarized SP modes.

The second medium is  modeled~\cite{Mai07,EZ06}
by complex dielectric permittivity and magnetic permeability
\begin{align}
\label{eq:eps2mu2}
	\varepsilon_2(\omega)=\varepsilon_{b}-\frac{\omega_\text{e}^2}{\omega(\omega+\text{i}\gamma_\text{e})},\\
	\mu_2(\omega)=\mu_{b}-\frac{F \omega^2}{\omega(\omega+\text{i}\gamma_\text{m})-\omega_{r}^2}
\end{align}
where~$\omega_\text{e}$ is the electron plasma frequency,
which is usually in the ultraviolet region, and $\gamma_\text{e}$ ($\gamma_\text{m}$) is the electric (magnetic) part damping rate.
F and $\omega_\text{r}$ are constants that depend on the geometry of the NIMM system. Thus the SP wave number is complex, its real part
gives the SP dispersions while the imaginary part gives losses due to the damping terms.

\begin{figure}[htp]
  \centering
  \includegraphics[width=80mm]{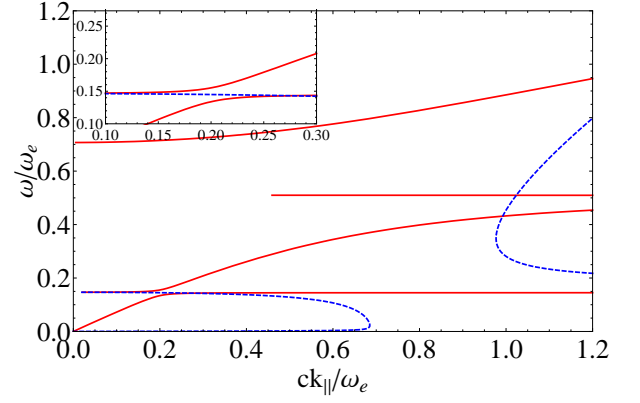}
  \caption{Dispersion curves for surface polaritons. The solid line is for electric TM modes, while dashed
  one refers to magnetic TE modes, for the set of
parameters ~$\omega_\text{e}=1.37\times 10^{16}\text{s}^{-1}$, $\gamma_\text{e}=2.73\times 10^{13}\text{s}^{-1}$,
$\gamma_\text{m}=\gamma_\text{e}/1000$, $\omega_\text{r}=0.145\omega_\text{e}$, and $F=0.6$. $\varepsilon_{1}=1.85$, $\mu_{1}=1$,
and we fix the background permittivity and permeability at $\varepsilon_{b}=2$, $\mu_{b}=2.5$.}
\label{fig:disps}
\end{figure}

\begin{figure}[h]
  \centering
 \includegraphics[width=80mm]{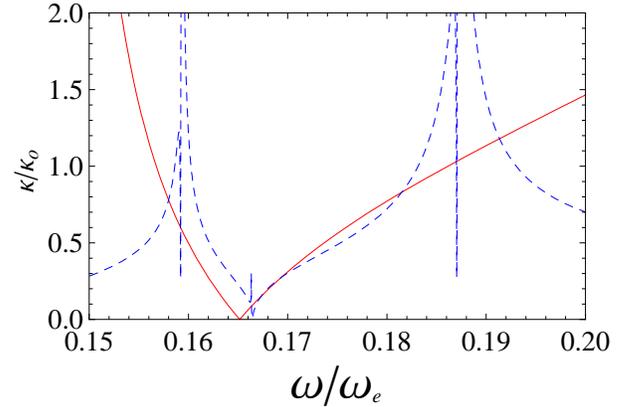}
 \caption{(Color online) Absorption loss for surface polaritons as a function of frequency $\omega/ \omega_\text{e}$ for TM ( solid line ) and TE ( dotted line).Losses are suppressed near frequency $\omega_0=0.167\omega_{e}$. $\kappa_o\approx 10^3/m $ for TM modes and $10^5/m $ for TE case.}
\label{fig:losses}
\end{figure}

Figure~(\ref{fig:disps}) shows the dispersion relations for both TM (solid line) and TE (dotted line) for the set of
parameters ~$\omega_\text{e}=1.37\times 10^{16}\text{s}^{-1}$, $\gamma_\text{e}=2.73\times 10^{13}\text{s}^{-1}$,
$\gamma_\text{m}=\omega_\text{e}/1000$, $\omega_\text{r}=0.145\omega_\text{e}$, and $F=0.6$. $\varepsilon_{1}=1.85$, $\mu_{1}=1$,
and we fix the background permittivity and permeability at $\varepsilon_{b}=2$, $\mu_{b}=2.5$. 
The inset shows the frequency range of 
interest between $0.1-0.3 \omega_\text{e}$. Near the frequency $0.14\omega_\text{e}$, there is a narrow gap. We shall be interested in 
the frequency $0.166\omega_\text{e}$ which lies above 
the gap region.
The imaginary part of equation ~(\ref{eq:TM-SP-existence-b}), $\kappa(\omega)$, yields SP loss
which is shown in Fig.~\ref{fig:losses} for both TM and TE polarized SP fields.The results reveal dips for the frequency~$\omega_0$
where losses are suppressed i.e. $\kappa(\omega_0)\sim0$. For later use we mention here that the frequency $\omega_0=0.16\omega_e$ will correspond or close to the transition wavelength in the DEIT medium of interest i.e. Pr:YSiO which is 606 nm.
We call the low loss surface plasmon modes as LLSP modes. Conventional surface plasmons at a metal interface support TM modes only, such modes are known to be lossy. In the presence of a meta-material interface we have seen that surface polaritons exist in both TM and TE polariaztion. Furthermore these modes have low losses and this is only possible with meta-material interface. These low loss surface polaritons with two polarization degrees of freedom constitute surface polariton qubits.

The electric fields~$E$ associated with these low loss modes are given for mode $\lambda=TM,TE$;
\begin{equation}
	\bm{E}_{\lambda}(\bm{r},t)=\bm{E}_{\text{o},\lambda}(K_\parallel,z)\text{e}^{\text{i}(K_\parallel x-\omega t)}, 
\end{equation}
where $\bm{E}_{\text{o},\lambda}(k_\parallel,z)$ is the SP field amplitude. In the low loss frequency range $K_\parallel\approx k_\parallel $ and the SP quantisation will determine the SP field amplitude which gives the SP coupling to atomic systems that we need to consider in later sections. 
We write the quantized SP field operator in the plane wave expansion as 
\begin{equation}
	\bm{E}(\bm{r},t) = \sum_\lambda \int\text{d}k_\parallel
		\left[\bm{E}_{\text{o}\lambda}(k_\parallel)a_\lambda(k_\parallel)
			\text{e}^{(\text{i}k_\parallel x-\omega t)} + \text{h.c.}\right],
\end{equation}
where the creation and annihilation operators obey the usual commutation relation
\begin{equation}
\label{eq:usualcommutation}
	\left[a_\lambda(k_\parallel),a^\dagger_{\lambda'}(k'_\parallel)\right]
		= 2\pi\delta_{\lambda\lambda'}\delta\left(k_\parallel-k'_\parallel\right).
\end{equation}
The field amplitude $\bm{E}_{0\lambda}$ is determined by matching the Hamiltonian in 
a dispersive ~\cite{KMS08}, but lossles medium 
\begin{align}
	H_\text{F}=\frac{1}{2}\int\text{d}^3r
		\left[\varepsilon_0\tilde{\varepsilon}\langle|E|^2\rangle
                +\mu_0\tilde{\mu}\langle|H|^2\rangle\right].  
\end{align}
with the quantized Hamiltonian
\begin{equation}
	H_\text{F}=\frac{1}{2}\int\text{d}k_\parallel\hbar\omega(k_\parallel)
		\left[a^\dagger(k_\parallel)a(k_\parallel)+a(k_\parallel)a^\dagger(k_\parallel)\right].
\end{equation}
where $\tilde{f}=\partial(f\omega)/\partial\omega$, and the $\langle...\rangle$ indicates the time average.
Using the equations above, we obtain the field amplitude ( for the TM case ) in the form
\begin{align}
	\bm{E}_\text{0}(k_\parallel,z)
		= N(k_\parallel)
		\theta(z)\left(\hat{x}+\text{i}\frac{k_\parallel}{k_1}\hat{z}\right)\text{e}^{-k_1z},\nonumber\\
                +N(k_\parallel)\theta(-z)\left(\hat{x}-\text{i}\frac{k_\parallel}{k_2}\hat{z}\right)\text{e}^{k_2z}.
\end{align}
where $\theta(z)$ is the well known step function. The normalization factor~$N(k_\parallel)$ determines the SP field amplitude and is given by
\begin{align}
N(k_\parallel)=&\sqrt{\frac{\hbar\omega(k_\parallel)}{2\pi\varepsilon_{o} L_{y} L_{z}}},
\end{align}
\begin{align}
\label{eq:LDS-TM}
        L_{z}
		=&\left[D+\frac{\omega^2(k_\parallel)}{c^2}S\right],\\
	D=&\tilde \varepsilon_1 \frac{|k_1|^2+|k_\parallel|^2}{|k_1|^2}\zeta_1
		+\tilde \varepsilon_2 \frac{|k_2|^2+|k_\parallel|^2}{|k_2|^2}\zeta_2,\\
	S=&\tilde \mu_1	\frac{|\varepsilon_1|^2}{|k_1|^2}\zeta_1
		+\tilde \mu_2 \frac{|\varepsilon_2|^2}{|k_2|^2}\zeta_2
\end{align}

and where the field confinement is defined as
\begin{align}
\label{eq:LDS-TM}
      	\zeta_{j}= \frac{1}{\text{Re} [k_{j}]} ,  j=1,2.
\end{align}
 The quantity $L_y$ is the medium size along $y$-direction, and $L_{z}$ is the mode length along the $z$-axis which is a function of the field confinement that is determined by the physical properties of the two media, viz-a-viz permittivities~$\varepsilon_{1,2}(\omega)$ and permeabilities~$\mu_{1,2}(\omega)$.
We emphasize that these equations are general and can be applied to any material with arbitrary set of parameters appropriate for surface polaritons.
Moreover, appropriate choice of materials and of frequencies of SP modes
i.e.\ adjusting the pairs ~$\varepsilon_{1,2}(\omega)$ and ~$\mu_{1,2}(\omega)$ can  lead to large confinement ( small $\zeta_{j}$ ) and thus decreases $L_z$ hence providing considerable enhancement of the SP field amplitude. This property will be used to increase the interaction coupling between the LLSP fields and atomic ensemble.

\section{Surface polariton control}
\label{sec:control}

 To control SP fields we use EIT approach, and study the dynamics of interaction between SP probe field and an ensemble of a lambda type three level atoms in medium~1 near the interface with a NIMM medium. The atom has two lower levels 1 and 2, and an upper level 3. We assume a control field acting on the atomic transition 2-3 while a probe field with frequency close to the frequency of atomic transition 1-3, and the transition 1-2 is dipole forbidden. The probe field is taken to be TM-polarized SP, while the control field need not be specified at the moment. The control and probe fields could be of the same polarization or they could be different.
The total Hamiltonian of the atom-SP field is
\begin{equation}
	\hat{H}=\hat{H}_\text{A}+\hat{H}_\text{F}+\hat{H}_\text{int}
\end{equation}
where free atom Hamiltonian~$\hat{H}_\text{A}$ is
\begin{equation}
\label{eq:HA}
	\hat{H}_\text{A}=\sum_j\hbar\omega_{31}^j P_{33}^j+\hbar\omega_{21}^j P_{22}^j,\,
	       \hbar\omega_1=0,
\end{equation}
for~$P_{mn}^j=|m\rangle _{jj}\langle n|$.
The interaction Hamiltonian is given in the dipole approximation as
\begin{equation}
\label{eq:HA}
	\hat{H}_\text{int}=-\sum_j \bm{d}_j\cdot \bm{E}(\bm{r}_j),
\end{equation}
where $\bm{E}(\bm{r}_j,t)$ is the SP field given above at the position $\bm{r}_j$, and the summation above runs over all the atoms in the interaction volume.

\begin{figure}[htp]
  \centering
  \includegraphics[width=80mm]{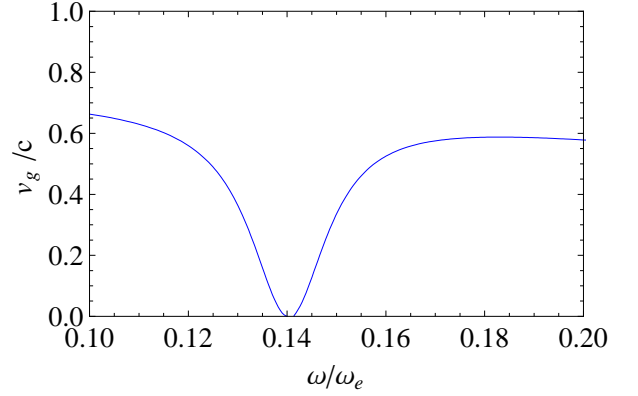}
  \caption{Surface polariton group velocity (in absence of interaction with three level atoms) in units c, as a function $\omega/\omega_\text{e}$ for TM.}
\label{fig:vgroup}
\end{figure}

Using the Heisenberg picture of motion we derive the operator equations in the limit of weak SP fields;
\begin{align}
\label{eq:Hpicture}
\frac{\partial}{\partial t}\hat{a}(k_\parallel,t)
	=&-\left[\text{i}\omega(k_\parallel)+\kappa(\omega_{31})\right]\hat{a}(k_\parallel,t)
		\nonumber	\\	&
		+ \text{i}\sum_jg\text{e}^{-k_1^\text{s}z^j}\text{e}^{-\text{i}k_\parallel x^j}P_{13}^j,
\end{align}
for $g=\bm{d}_{31}\cdot \bm{E}_0^*/\hbar$ the strength of SP dipole coupling to atom~$j$
and summation taken over all atoms~$\{j\}$ close to the interface.
Atomic operators $P_{mn}^j=|m\rangle_{jj}\langle n|$ satisfy
\begin{align}
\label{eq:PP}
\frac{\partial}{\partial t}P_{13}^j
	=&-\left(\text{i}\omega_{31}^j+\gamma_{31}^j \right)P_{13}^j+\text{i}\Omega_\text{c}(r_j)P_{12}^j
			\nonumber	\\	&
		+\text{i}\int\text{d}k_\parallel\left(\bm{d}_{31}\cdot \bm{E}_0/\hbar\right)
		\text{e}^{-k_1^\text{s}z^j}\text{e}^{\text{i}k_\parallel x^j}
		\hat{a}(k_\parallel,t),
			\nonumber	\\
\frac{\partial}{\partial t}P_{12}^j
	=&-\left(\text{i}\omega_{21}^j+\gamma_{21}^j \right)P_{12}^j
		+\text{i}\Omega_\text{c}^*(\bm{r_j})P_{13}^j,
\end{align}
where $\Omega_\text{c}(\bm{r}_j)=(\bm{d_{32}}\cdot \bm{E}_0^c/\hbar)$ is Rabi frequency of the interaction with control SP field. The index s refers to the probe field while c to control field. The probe and control wave numbers along z-direction
\begin{equation}
	k_1^{s,c}=\sqrt{k_\parallel^2(\omega_{s,c})-{\omega_{s,c}}^2 \varepsilon_1(\omega_{s,c})\mu_1(\omega_{s,c})/c^2},
\end{equation}
are calculated at the resonant frequencies $\omega_{31}$ and $\omega_{32}$, respectively.
The transition frequencies are related to the wave number $k_\parallel$ by the dispersion relation appropriate for the type of polarization. The phenomenological decay constants $\gamma_{21}$ and $\gamma_{31}$ are determined by the interactions of jth atom with other field modes and environment, and  the cavity loss is given by $\kappa(\omega_{31})$ as discussed in previous section. Adopting the usual EIT approximations~\cite{Mil05,FIM05}, and performing algebraic calculations we arrive at an equation for the slowly varying SP probe field
$ \hat A(t,x)=\text{e}^{-\text{i}k_\parallel^s(\omega_{31}) x}\int\text{d}k_\parallel\bm{E}_0(k_\parallel)
\hat a(k_\parallel,t)\text{e}^{\text{i}k_\parallel^s x}$,
and using the Fourier transformation $ \hat{A}(\nu,x)=(2\pi)^{-1}\int\text{d}t\text{e}^{\text{i}\nu t} \hat{A}(t,x)$ we find for field A the expectation value $\langle \hat{A}(\nu,x) \rangle =A(\nu,x)$:
\begin{align}
\label{eq:A}
	\Bigg(\frac{\partial}{\partial x}-\text{i}\frac{\nu}{v_0}\Bigg) A(\nu,x)
		=-\left[\alpha(\nu)+\kappa(\omega_{31})\right] A(\nu,x),
			\nonumber	\\
	A(t,x)=\int_{-\infty}^\infty\text{d}\nu\,
		e^{-\text{i}\nu t+[\text{i}\frac{\nu}{v_0}
		-\alpha(\nu)-\kappa(\omega_{31})]x}
		A(\nu,0),
\end{align}

\noindent
where $\nu=\omega(k_\parallel)-\omega_{31}$
is the SP probe field detuning from the central frequency~$\omega_{31}$; the complex absorption coefficient $\alpha(\nu)$ is give by the expression:
 \begin{align}
\label{eq:alpha}
	\alpha=&\frac{2\pi|g|^2}{v_0(\omega_{31})}(\gamma_{21}-\text{i}\nu)
		\nonumber	\\	&
		\times\int_0^\infty\!\!\int_0^{L_y}\text{d}y\text{d}z\,
		\frac{n(\bm{r})\text{e}^{-2k_1^\text{s} z}}
		{\left|\Omega_\text{c}(\bm{r})\right|^2-\left(\nu+\text{i}\gamma_{21}\right)
			\left(\nu+\text{i}\Gamma_{31}\right)}.
\end{align}
The basic group velocity $v_0=v_0(\omega_{31})
	=\frac{\partial\omega}{\partial k_\parallel}\big|_{\omega_{31}}$ of the probe SP field (without interaction with three level medium) and the field amplitude $\bm{E_0}$ are calculated at the central frequency $\omega_{31}$ and $\Delta^j=\omega_{31}^j-\omega_{31}$ is the jth atomic detuning from the central frequency $\omega_{31}$.\\ Spectral behavior of group velocity $v_0$ is presented in Fig~(\ref{fig:vgroup}) for TM polarized SP fields.
We note that group velocity is close to zero near the gap edges around $0.14 \omega_{e}$.
In this analysis we work in a frequency range where SP losses are low
close to $\omega_0$ , so we take the central frequency $\omega_{31} \approx \omega_0=0.167\omega_e$ which corresponds to basic group velocity
of $v_\text{g}=0.6c$.

The control field, which yields~$\Omega_\text{c}$, can be a freely propagating mode or an SP TE or TM field.
Here we restrict to the latter case: $|\Omega_\text{c}(\bm{r})|^2=|\Omega|^2\text{e}^{-2k_1^c z}$
with~$k_1^\text{c}$ the control field wave number in the $z$-direction for medium~1.

As seen in Eq~(\ref{eq:alpha}) the complex absorption coefficient $\alpha(\nu)$ is modulated by the spatial exponential factors so that the basic spectral properties of the SP-SP control are sensitive to the relative spatial behavior of the modes. In Eq~(\ref{eq:alpha}) we have taken into account the inhomogeneous broadening of the resonant line by assuming the weight function $h(\Delta)=(\Delta_w/\pi)/(\Delta^2+\Delta_w^2)$ (where $\Delta_w$ is an inhomogeneous broadening width, and $\Delta\approx \Delta^j$);$\Gamma_{31}=\Delta_w+\gamma_{31}=\eta\gamma_{31}^{sp}$, $\gamma_{31}^{sp}=\frac{\omega_{31}^3|d_{31}|^2}{3\pi\varepsilon_0\hbar c^3}$ is the spontaneous decay rate in free space, and $\eta$ is a factor that takes into account additional contribution to the line width due to the other processes e.g. inhomogeneous broadening, collision etc. We assume for the optical transitions the following typical values $\gamma_{31}^{sp}\approx10^3 s^{-1}$ and $\eta\approx10^6-10^7$ in some laser crystals such as ruby at Helium temperature~\cite{Mil05}. The real part of $\alpha(\nu)$ gives the absorption coefficient and the imaginary part gives the dispersions. In general the atomic density distribution~$n(y,z)$ depends on both y and z directions, but here we shall consider only the situation where~$n(y,z)=n_0(z)$  i.e. depends on the distance from the interface.

The atomic density~$n_0(z)$ can vary with distance z from the interface depending on the preparation of the atomic ensemble that is an important aspect for experiment. In particular we mention that optical pumping by the resonant laser fields can be used for preparing the atoms in the initial quantum state. As seen in Eq~(\ref{eq:alpha}) the coupling of the atoms with probe and control fields are strong for atoms with spatial coordinates close to the interface. It is reasonable to use the atomic ensemble prepared within a small layer close to the interface with thickness~$z_0$ such that~$n_0(z)=n_0$ for ~$0<z<z_0$ where ~$n_0$ is constant. It is important in this analysis also to investigate the optimal value for~$z_0$ where the SP control can be effectively realized. We can obtain the absorption coefficient in this case by integration over z in Eq~(\ref{eq:alpha}) in a closed form in terms of
the hypergeometric function as follows;
\begin{equation}
\label{eq:alphanuz0}
	\alpha(\nu,z_0)=\alpha_0(\omega_{31})G\left(k_1^\text{s},k_1^\text{c},z_0,\beta(\nu)\right),
\end{equation}
where
\begin{equation}
	\alpha_0(\omega_{31})=\frac{n_0L_y}{2k_1^\text{s}} \frac{2\pi|g|^2}{v_0(\omega_{31})\Gamma_{31}}
\end{equation}
is the absorption coefficient of the atoms at~$\omega_{31}$ in the absence of the control field for~$z_0>1/k_1^\text{p}$, and
\begin{align}
\label{eq:G}
	G(k_1^\text{s},k_1^\text{c},z_0,\beta)
		=\frac{\text{i}\Gamma_{31}}{\nu+\text{i}\Gamma_{31}}
		\Big\{  {_2F_1}\left(1,\frac{k_1^\text{s}}{k_1^\text{c}},\frac{k_1^\text{s}+k_1^\text{c}}{k_1^\text{c}},
			\frac{1}{\beta(\nu)}\right)
					\nonumber \\	
		-\text{e}^{-2k_1^\text{s}z_0}\; _2F_1\left(1,\frac{k_1^\text{s}}{k_1^\text{c}},
			\frac{k_1^\text{s}+k_1^\text{c}}{k_1^\text{c}},
			\frac{\text{e}^{-2k_1^\text{c}z_0}}{\beta(\nu)}\right)\Big\},
\end{align}
where ${_2F_1}$ is the well known hyper geometric function, and
$\beta
		=(\nu+\text{i}\gamma_{21})
		(\nu+\text{i}\Gamma_{31})/|\Omega_c|^2$.
The function $G(k_1^\text{s},k_1^\text{c},z_0,\beta)$ characterizes the spectral properties of the SP control and takes numerically maximum value of 1.

\section{Spectral control and quantum memory of SP}
\label{sec:spectral}

In the presence of the control field~$\Omega_\text{c}$, the main spectral behavior of $\alpha(\nu,z_0)$
is sensitive to the spatial confinement of SP fields expressed via the hyper geometric functions in Eq~(\ref{eq:G}). The general solution in Eq~(\ref{eq:G}) has a rich spectral behavior and depends on the ratio between the control and probe fields wave numbers, but we consider few simple cases which could be very close for practical control of slow SP field in the limit of very large thickness of atomic layer i.e.
$z_0 \gg (2k_1^\text{s})^{-1}$, where it is possible to reduce the general hyper geometric solutions to more tractable forms. It is quite interesting that the spatial heterogeneity of the interaction determines interesting new spectral properties of the absorption coefficient in different regimes of SP-control as we discuss in the following cases of interest that are analyzed at the resonant frequency $\nu \approx0$;
\begin{align}
\text{(i)}\frac{k_1^s}{k_1^c}\gg1; 
         G\approx\frac{\text{i}\Gamma_{31}}{\nu+\text{i}\Gamma_{31}}\frac{\beta(\nu)}{\beta(\nu)-1}
         (1-\frac{k_1^c/k_1^s}{\beta(\nu)-1})|_{\nu\approx0}\rightarrow0.
\end{align}
For $k_1^c/k_1^s=0$, this gives the usual spectral properties of the well-known EIT.
\begin{align}
\text{(ii)}\frac{k_1^s}{k_1^c}=\frac{1}{2}; 
       G=\frac{\text{i}\Gamma_{31}}{\nu+\text{i}\Gamma_{31}}\sqrt{\beta(\nu)}\text{arctanh}[\sqrt{1/\beta(\nu)}|_{\nu=0} \nonumber\\
       \rightarrow\frac{\text{i}\pi}{2}\sqrt{\beta(\nu)}\rightarrow0.
\end{align}
\begin{align}
\text{(iii)}\frac{k_1^s}{k_1^c}=1; 
         G=\frac{\text{i}\Gamma_{31}}{\nu+\text{i}\Gamma_{31}}\beta(\nu)\text{ln}[1-\frac{1}{\beta(\nu)-1}|_{\nu=0} \nonumber\\
         \rightarrow\beta(\nu)\text{ln}[\beta(\nu)]_{\nu=0}\rightarrow0.
\end{align}
This case , $k_1^s/k_1^c=1$, is most convenient for experimental realization of SP control.
\begin{align}
\text{(iv)}\frac{k_1^s}{k_1^c}=2; 
         G=\frac{-2\text{i}\Gamma_{31}}{\nu+\text{i}\Gamma_{31}}\beta(\nu)(1+\beta(\nu)\text{ln}[1-\frac{1}{\beta(\nu)}]|_{\nu\approx0} \nonumber\\
       \rightarrow-2\frac{\beta(\nu)}{\beta(\nu)-1}(1-\frac{k_1^s/k_1^p}{\beta(\nu)-1})|_{\nu\approx0}\rightarrow0.
\end{align}
 
\begin{figure}[h]
  \centering
  \includegraphics[width=80mm]{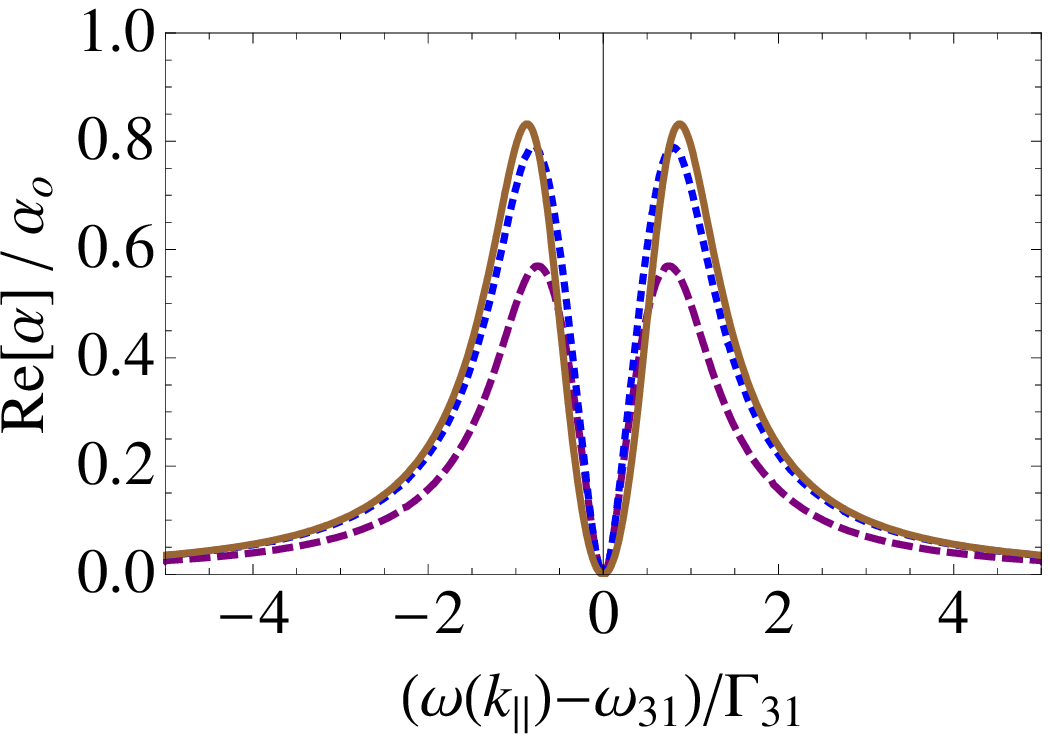}
  \includegraphics[width=80mm]{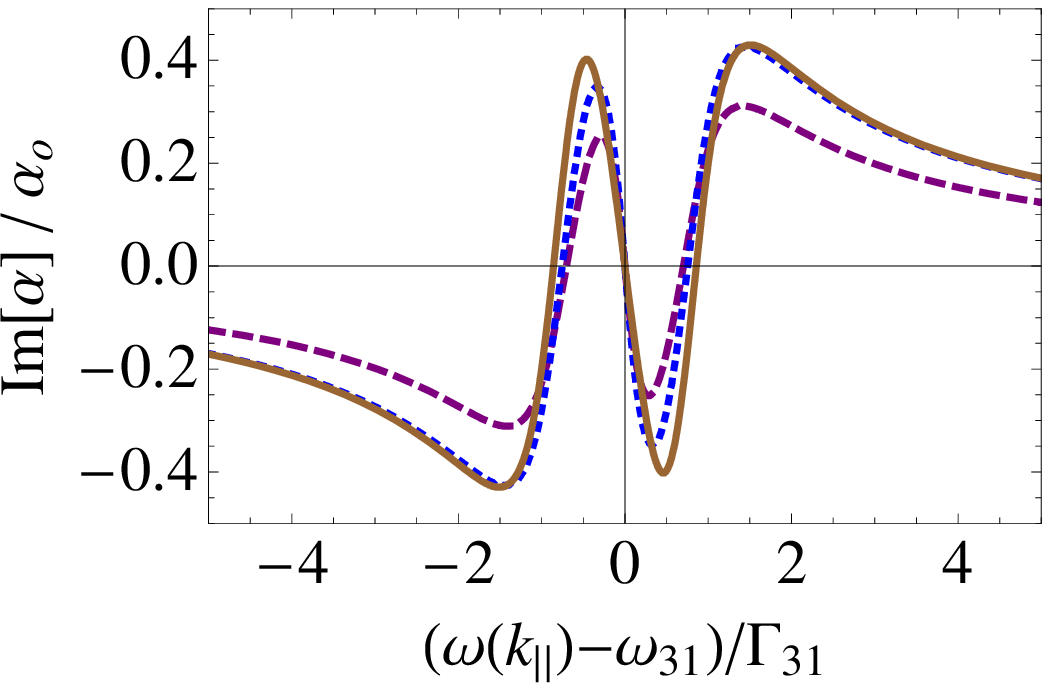}
  \caption{(a) The absorption coefficient ( real part of $\alpha(\nu,z_0)$) and (b) the dispersion ( imaginary part of $\alpha(\nu,z_0)$ as functions of SP detuning $\nu=\omega(k_\parallel)-\omega_{31}$, $\Gamma_{31}=10^9\text{s}^{-1}$ for fixed $z_0=1/k_1^s$ and different $k_1^s/k_1^c$ ratios; red (dashed) at $k_1^s/k_1^c=1/2$, blue (dotted) at $k_1^s/k_1^c=1$ , and brown (solid) at $k_1^s/k_1^c=2$.}
\label{fig:abs1}
\end{figure}

\begin{figure}[htp]
  \centering
  \includegraphics[width=80mm]{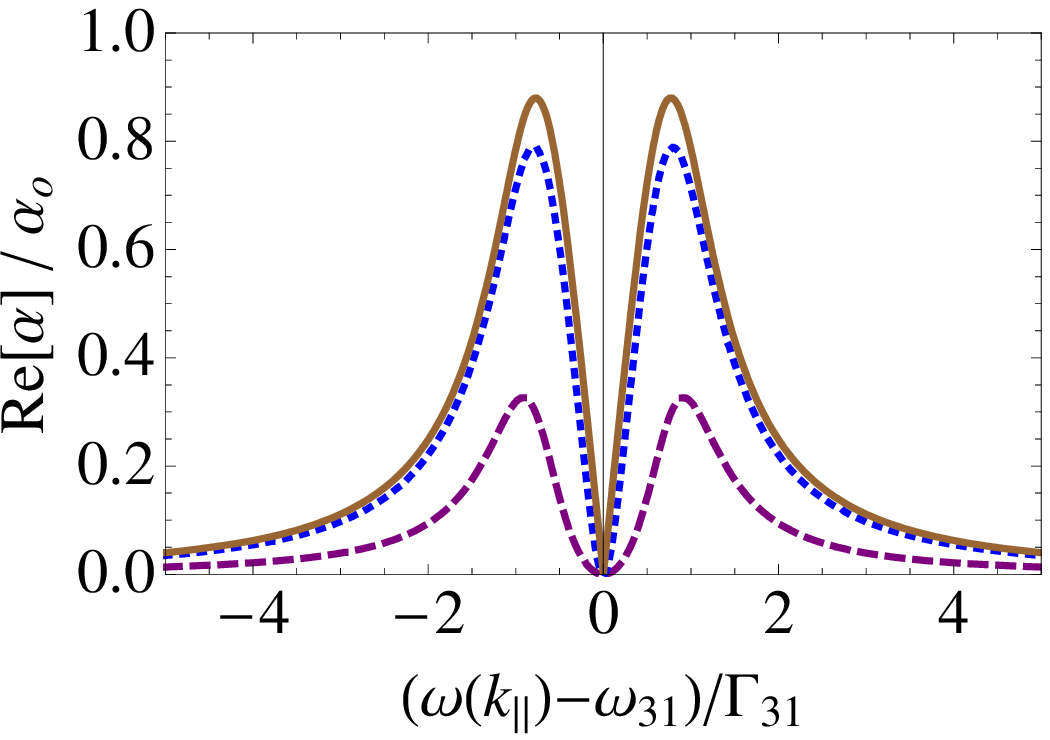}
  \includegraphics[width=80mm]{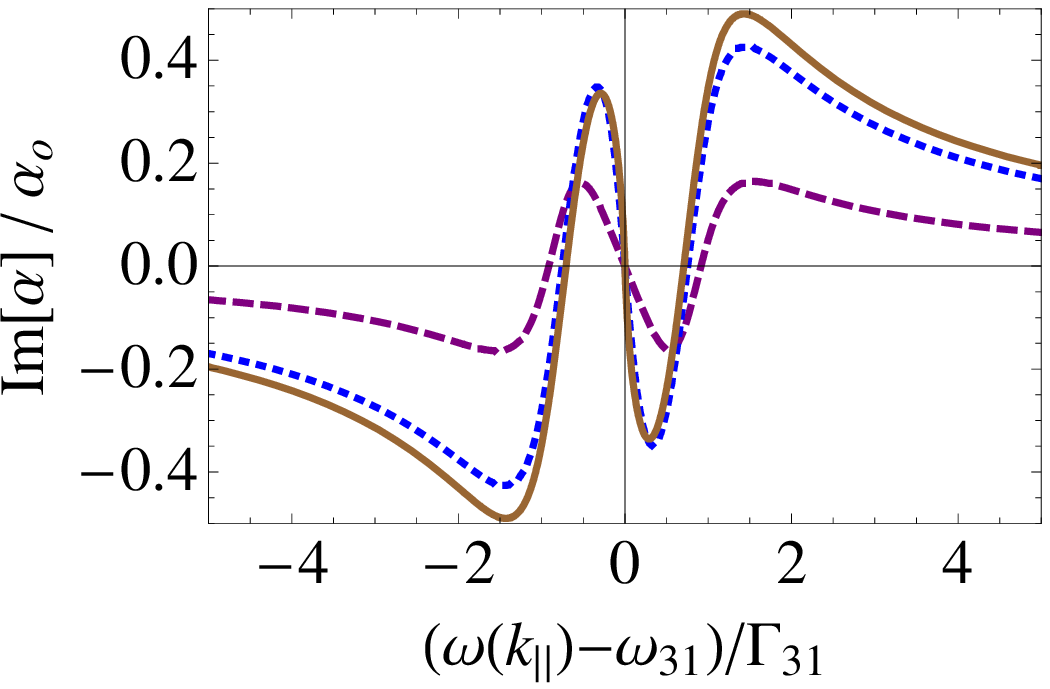}
  \caption{(a) The absorption coefficient ( real part of $\alpha(\nu,z_0)$) and (b) the dispersion ( imaginary part of $\alpha(\nu,z_0)$ as functions of SP detuning $\nu=\omega(k_\parallel)-\omega_{31}$, $\Gamma_{31}=10^9\text{s}^{-1}$ for fixed $k_1^s=k_1^c$ and different atomic layer thickness $z_0$ ; red (dashed) at $z_0=0.5/k_1^s$, blue (dotted) at $z_0=1/k_1^s$ , and brown (solid) at $z_0=10/k_1^s$.}
\label{fig:abs2}
\end{figure}

 As it is seen from the above discussions that the absorption of the probe SP field is highly suppressed for all cases in the spectral range around $\nu \approx0$.
We take the EIT medium as Pr:YSiO ( see next section for discussion ), assuming the following parameters ~$n_0\approx10^{20}\text{cm}^{-3}$, $L_y\approx 10\lambda $ (where $\lambda=606nm $ is the transition wavelength of our EIT medium ), $z_0\approx1/k_1^s\approx 1 \mu\text{m}$,
and $\eta=10^6$, we estimate ~$1/\alpha_0(\omega_{31})=1 \mu\text{m}$. The curves in Fig~(\ref{fig:abs1}) and Fig(~\ref{fig:abs2}) show the absorption and dispersion profiles ( in units of $1/\alpha_0(\omega_{31})= 1 \mu\text{m}$) for the cases (ii),(iii), and (iv) above.
We note also that the cases (iii) and (iv) of the SP control have more sharp spectral shape due to the fact that the atoms at distance $z_0\gg(k_1^s)^{-1}$ make a large contribution to the SP control at the condition of weak control field amplitude. 
The described four  cases  demonstrate an interesting possibility to control the spectral properties of window transparency for LLSP field by appropriate choice of the atomic thickness $z_o$. Furthermore we note that the spectral properties of the window transparency can be varied even more extensively by using another spatial shapes of the atomic concentration along $z-$ direction.

\begin{figure}[htp]
  \centering
  \includegraphics[width=80mm]{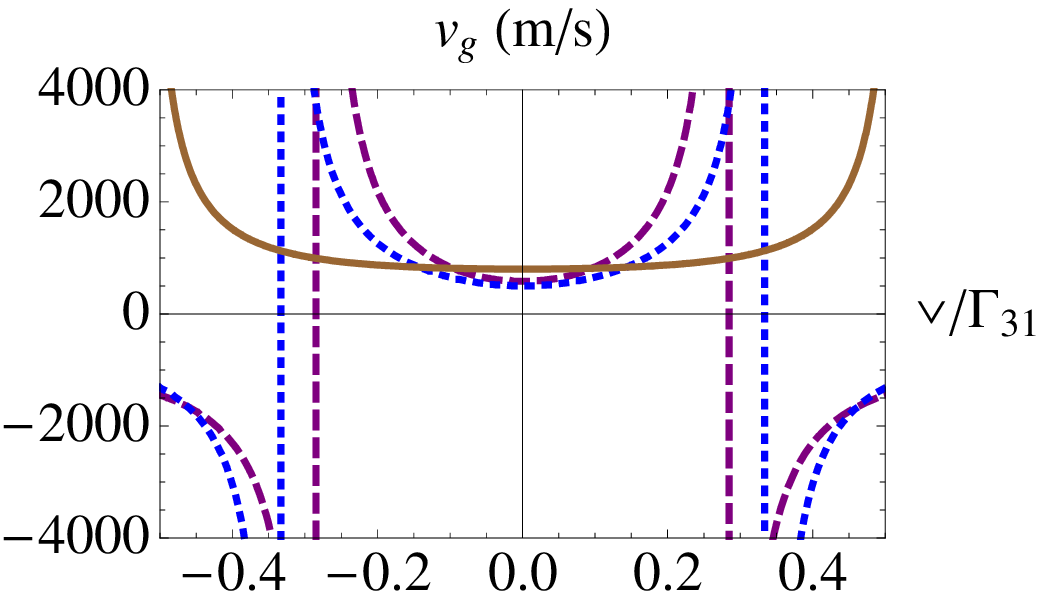}
  \includegraphics[width=80mm]{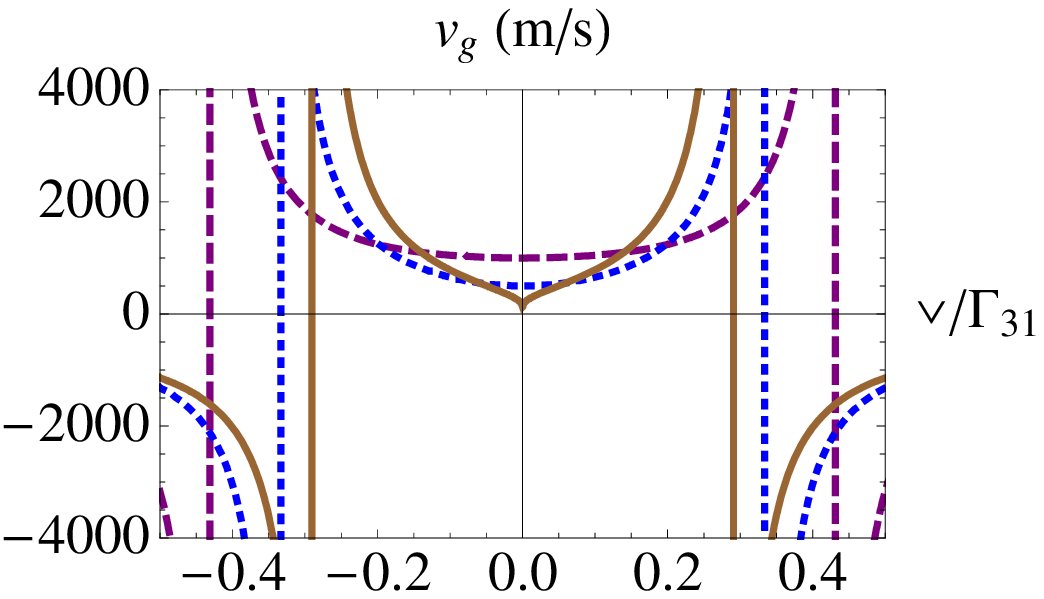}
  \caption{The SP group velocity due to the interaction with three level atom ensemle as functions of $\nu=\omega(k_\parallel)-\omega_{31}$ (a) for fixed $z_0=1/k_1^s$ and different ratios $k_1^s/k_1^c$: red (dashed) at $k_1^s/k_1^c=1/2$ , blue (dotted) at $k_1^s/k_1^c=1$ ,and brown (solid) at $k_1^s/k_1^c=5$, and (b) for the fixed ratio $k_1^s/k_1^c=1$ and different atomic layer thicknesses: red (dashed) at $z_0=0.5/k_1^s$, blue (dotted) at $z_0=1/k_1^s$, and brown (solid) at $z_0=10/k_1^s$.}
\label{fig:vg}
\end{figure}

Spectral properties of the group velocity play an important role in SP field control. Using Eq~(\ref{eq:A}) we find the group velocity of the probe SP field near $\omega (k_{\parallel}) \approx \omega_{31} $
\begin{equation}
\label{eq:VG}
v_g(\nu)=\frac{v_0(\omega_{31})}{1-\alpha_0 v_0(\omega_{31})\text{Im}[\frac{\partial}{\partial \nu}G(k_1^\text{s},k_1^\text{c},z_0,\nu,\beta)]},
\end{equation}
where the derivative $\frac{\partial}{\partial\nu}G$ is expressed through the same hyper geometric function. The group velocity behavior depends on the physical parameters $\alpha_0,\omega_{31},k_1^p/k_1^s,z_0,\nu,\Omega_c$ and because of lack of space we only discuss the most important properties for control of SP field propagation. Spectral behavior of the group velocity $v_g(\nu)$ as in Eq~(\ref{eq:VG}) close to the exact two-photon resonance is shown in Fig~(\ref{fig:vg}) that demonstrates large enough reduction from $v_0(\omega_{31})\approx 0.6c$ (in the absence of interaction with three level atom ensemble) to $v_{g}(\nu\approx0)\approx500\text{m/s}$ (due to interaction with three level medaium) at moderate parameters of the atomic systems 
$(n_0=10^{20}\text{cm}^{-3})$ and large inhomogeneous broadening $\Gamma_{31}=10^9\text{s}^{-1}$ .

For the pulse duration $\delta t= 0.1 \mu \text{sec} $, the reduced group velocity leads to longitudinal pulse compression $l_{SP}=v_g\delta t=50\mu\text{m}$ that is 20 times smaller than the propagation length $( L=1/\kappa(\omega_{31}) \approx 1 mm)$ of SP fields. Thus the SP-field pulse can be successfully stored in the long-lived atomic coherence $\rho_{21} $ and retrieved by switching the control field $\Omega_\text{c}$~\cite{FL02}. The demonstration of the possibility of such spectral manipulation and quantum memory of the weak LLSP field is important for quantum informatics with nanoscale plasmonics.The proposed manipulation can be useful for single photon and for more intensive light fields as well. 
As seen in Fig. 3, simultaneous excitation of the LLSP modes with TM and TE polarization in the same frequency range  close $\omega_o$ gives rise to quantum manipulation of the polarization qubits carried by the LLSP field. In particular one can store the polarization qubit in nanoscaled quantum memory by using the above described scheme taking into account additional polarization properties of the atomic transitions. It is reasonable to find a possibility of most nontrivial manipulation of the LLSP fields in nonlinear interactions of the TM and TE LLSP modes due to higher enhancement of the LLSP field near the dielectic/NIMM interface. Below we describe one of such nonlinear interaction for two temporally selected LLSP modes.

\section{Nonlinear quantum gates with SP modes}
\label{sec:nonlinear}

 In this section we show that weak LLSP fields can be used to generate large enhancement of Kerr nonlinear interaction between two LLSP pulses characterized by their different group velocities. For simplicity in the case of temporally separated LLSP mode we drop out possible manipulations of additional polarization degree of freedom which however can be a serious subject of further investigations.
The quest for large Kerr nonlinear coefficient and large phase shifts has been vigorously pursued by many
authors \cite{SI96,HH00,LI01,PK02,KZ03,Ott03,And05,Ch06,WMS06,WANG08,MCL08,PF08} over the past years and promising success has been reported. One of the primary interest in large Kerr nonlinearity in quantum information, is that large nonlinearity enables the implementation of some quantum gates , e.g. controlled phase gate. Our scheme here applies to a gas system like rubidium atoms, as well as to solid state system like Pr:YSiO. We focus here on the a solid system for the following reasons; 1) EIT has been experimentally realized in this system by many authors \cite{ALSM06,HHS97,TSSMHH02}, and it could realize double EIT effects \cite{KZ03,WANG08,MCL08} that are the mechanism we adopt in this paper to achieve large phase shifts, 2) this system is useful for information storage and retrieval and storage time of few seconds has already been experimentally reported \cite{ALSM06}, 3) being solid state this system could avoid many problems that are present in gaseous systems e.g Doppler broadening due to the fact that atoms are locked into the solid with limited movement, and finally 4) the wavelength of this system (like the gas system) is commensurate with the current NIMM technology that has reached wavelength of 580 nm \cite{Sha09}.

We focus our attention only on one interesting scheme of the nonlinear interaction \cite{WMS06,MCL08} which uses 5-level atomic scheme (5LA) in ${\rm Rb}^{87}$ that leads to uniform ~\cite{R93} nonlinear phase shift for two slowly propagating interacting light pulses at the condition of double EIT effect. We adopt this 5LA scheme for our SP fields in the solid system Pr:YSiO. We investigate the possibility to enhance the nonlinear interaction between weak light fields by creating slowly co-propagating LLSP fields characterized by large transverse and longitudinal confinement and by increasing the interaction time  due to the EIT condition.

\begin{figure}[h]
 \centering
\includegraphics[width=40mm]{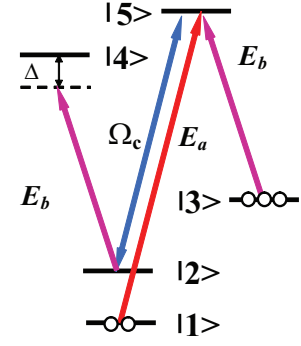}
\caption{Frequencies and fields of two slow LLSP fields $a,b$, of the control LLSP field $\Omega_c$ and energy diagram of rare earth Pr ions in YSiO crystal (Pr:YSiO).}
\label{fig:6LA}
\end{figure}

Let us assume that the two interacting LLSP pulses are excited one by one at the interface input with slightly different adjusted group velocities $\upsilon_{a,b}$. Let the second LLSP pulse have larger group velocity $\upsilon_b>\upsilon_a$ and outrace the first LLSP at the medium output as it is depicted in Fig\ref{fig:scheme}. Energy diagram of the 5LA system and level scheme of Pr:YSiO are shown in Fig.~\ref{fig:6LA}.
 As in previous works \cite{WMS06,MKS10,MWMS10} we derive nonlinear coupled equations for two slowly propagating LLSP fields by taking into account spatial confinement of interaction with resonant atomic systems in the spirit of \cite{KMS08}, and we get the Kerr nonlinear coefficient of field b due to cross phase modulation of field a as:
\begin{equation}
\label{nonlinear equation}
\left({\partial\over {\upsilon _b \partial t}} + {\partial \over {\partial x}}\right) E_{b}(t,x) = i\chi _{a} I_{a} (t,x) E_{b}(t,x),
\end{equation}

\begin{equation}
\label{nonlinear coefficient}
\chi _a = \frac{2\pi n_a z_{\rm 0}\Phi [(k_a^p + k_b^p - k^c)z_{\rm 0}]}{\hbar ^4\upsilon
_{b,o} \vert \Omega _c \vert ^2\Delta }
\left\langle {\vert \bm {d}_{24} \bm{E_b} \vert ^2\vert \bm{d}_{15} \bm
{E_a} \vert ^2} \right\rangle ,
\end{equation}

\noindent
where $I_a (t,x) =|E_a(t,x)|^2 $ for field a, and  $k_{a,b}^s$ is the real part of SP probe field (a or b) wave number along z and $k^c$ is the control field wave number along z. $\upsilon _\emph{l} = \upsilon _{\emph{l},o} / (1 + \beta _\emph{l} )$ is the group velocity of \emph{l}-th
slowly propagating LLSP pulse, $\upsilon _{\emph{l},o} $ is the group velocity of
\emph{l}-th LLSP pulse in absence of the resonant atoms, $\beta _b = 2\pi n_b z_{\rm 0}\Phi
[(k_b^p - k^c)z_{\rm 0}]\left\langle {\vert \bm {d}_{35} \bm E_b \vert ^2}
\right\rangle / (\hbar ^2\vert \Omega _c \vert ^2)$, $\Phi (y) = e^{ -
y}sinh(y) / y$, $z_{\rm 0}$ is spatial thickness of the atomic medium along z-direction,
$\Omega _c $ is the Rabi frequency of the
control field, $n_b$ is atomic density on the $3$-th level, $\Delta $ is the spectral detuning, $\bm
{d}_{24} , \bm {d}_{15} $ are atomic dipole moments of atomic transitions presented
in Fig.~\ref{fig:6LA}. $\bm{E}_\emph{l} $ ($l=a , b$) is the electric field of \emph{l}-th LLSP pulse which is given previously,
and $\left\langle {...}\right\rangle$ denotes here averaging over the orientation of the atomic dipole moments.
 The self phase modulation (SPM) of field b, $\chi _b $, is ignored in the above equation since this term can be compensated for interferometrically
as discussed in \cite{MKS10}, and references cited therein. The corresponding phase shift experienced by single quantum SP field b due to its nonlinear interaction with other single quantum LLSP pulse a as it traverses a medium of length $L$ is given (see also ~\cite{R93,MKS10,MWMS10}) as $\Phi_{\rm XPM}\cong\chi_{\rm a}(\omega_s)\times L/(\delta t  v_{{\rm a,0}})$ where $\delta t$ is the SP pulse temporal duration, and $\omega_s$ is the Pr:YSiO transition frequency of interest.

\begin{figure}[htp]
 \centering
 \includegraphics[width=80mm]{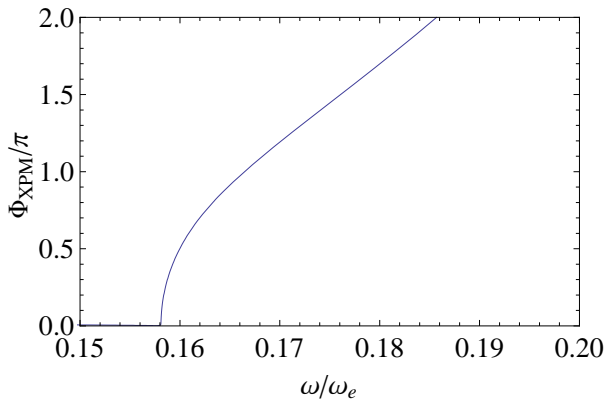}
 \caption{The phase shift due to SP cross phase modulation in double EIT scheme (in units $\pi$) as a function of probe field frequency $\omega/\omega_{\rm e}$.The frequency of interest is around $\omega \approx 0.167 \omega_{\rm e}$. }
\label{fig:Phase_shift}
\end{figure}

For the 5LA in Pr:YSiO we assume ideal EIT conditions which take place for small enough thickness of the atomic layer $z_{\rm 0}\approx1/k_j^p \approx 1/ k^c \approx1 \mu m$, the transition wavelenght is 606nm \cite{ALSM06,HHS97,TSSMHH02} the linewidth is about MHz and the detuning $\Delta=4.5$ MHz, Rabi frequency of control field $\Omega_c =1.5$ MHz. We estimate the dipole moments for such transition to be of the order $10^{-2}{\rm e}{\rm a}_{\rm 0}$ where $e$ is the electronic charge and ${\rm a}_{\rm 0}$ is the Bohr radius. We have chosen the media parameters such that the transition wavelength 606nm corresponds to SP frequency resonant with Pr:YSio transition frequency  $\omega_s=0.167\omega_{\rm e}$ which is close to the frequency where SP fields exhibit low losses and large confinement. The atomic density of levels is taken to be $10^{20} \text{cm}^{-3}$ that is close to typical solid medium, and the medium size is assumed $L \approx 0.3 \text{mm}$ and the SP pulse temporal duration is assumed of the order $\approx 0.1\mu s$. For these set of parameters which approximate the experimental data \cite{ALSM06,HHS97,TSSMHH02} we show in  Fig.~\ref{fig:Phase_shift} the phase shift of field b due to cross phase modulation with field a.
The results demonstrate clearly that it is possible to achieve all the requirements of: low losses, large confinement, large Kerr nonlinear coefficient and cross phase shift of order $\pi$. 
Finally we note that with the nonlinear $\pi$ phase shift, the LLSP pulses can be easily separated at the output from each other in real experiment if the input two LLSP pulses are characterized by different polarizations (i.e., for TM and TE LLSP pulses).
Whereas the analysis presented here pertains to the solid system Pr:YSiO, the low loss SP and their high confinement together with double EIT scheme is also applicable to gaseous system like ${\rm Rb}^{87}$ with appropriate choice of parameters \cite{MKS10}.

\section{Conclusions}
\label{sec:conclusions}

We have studied electric and magnetic surface polaritons in half space metamaterial in contact with a dielectric. We have demonstrated the possibility of slow low loss SP fields with small longitudinal spatial size of the SP pulse and highly reduced group velocity. Thus the proposed scheme has a large potential for the quantum control of the SP-fields, and applications in compact quantum memory devices~\cite{KP00,Moi07,KTGNKC06,ALSM06} localized near the surface of mate-materials.  We have discussed only few possibilities of the SP field control that indicate the rich physical parameters that can be used in order to realize the most convenient conditions that suit particular applications of this scheme. In particular it would be interesting to analyze  promising possibilities determined by the other types of geometry of the interface, as well as by using another spectral properties of the atomic systems and interaction with varied intensive control laser fields. Also we anticipate to find interesting possibilities for the SP field control for the atomic systems characterized by another spatial densities, in particular with periodic density modulation along the SP field propagation.

As we have found above, combining SP fields at NIMM interface with the DEIT mechanism
yields the trifecta for large cross-phase modulation . These three sought-after properties are
low loss, high field confinement, and large Kerr coefficients.
The goal is to reach mutual phase shifts of~$\pi$ at the single-photon level,
which would have profound implications for quantum information technology.

\acknowledgments
We gratefully acknowledge financial support from {\em i}CORE, NSERC, KACST,  RFBR grant \#08-07-00449 and 10-02-01348-a, Government contract of RosNauka 02.740.11.01.03, BCS is a CIFAR Fellow.

\end{document}